\documentclass[a4paper,12pt]{article}
\usepackage{graphicx,rotating,amsmath,amssymb,enumerate,array,braket}
\usepackage{hyperref}\usepackage{slashed,cite}
\pdfoutput=1
\usepackage{mathtools}
\hypersetup{colorlinks,bookmarksopen,bookmarksnumbered,
linkcolor=blus,pdfstartview=FitH,urlcolor=rossos,citecolor=verde}

\newcommand{\med}[1]{\langle #1 \rangle}

\newcommand{\fig}[1]{~\ref{fig:#1}}

\newcommand{\sfrac}[2]{{#1}/{#2}}

\newcommand{\SH}{S_{\rm hard}}
\newcommand{\Rop}{R}
\newcommand{\calW}{{\cal L}}

\allowdisplaybreaks
\usepackage{multicol}
\usepackage{color}
\definecolor{rosso}{cmyk}{0,1,1,0.4}
\definecolor{rossos}{cmyk}{0,1,1,0.55}
\definecolor{rossoc}{cmyk}{0,1,1,0.2}
\definecolor{blu}{cmyk}{1,1,0,0.3}
\definecolor{blus}{cmyk}{1,1,0,0.6}
\definecolor{bluc}{cmyk}{1,1,0,0.1}
\definecolor{verde}{cmyk}{0.92,0,0.59,0.25}
\definecolor{verdec}{cmyk}{0.92,0,0.59,0.15}
\definecolor{verdes}{cmyk}{0.92,0,0.59,0.4}
\usepackage[
  a4paper,
  width=17cm,
  height=22.5cm,
  centering, % This automatically makes left margin = right margin
  top=2cm    % Adjust top margin as needed
]{geometry}

\newcommand{\mub}{\bar{\mu}}
\renewcommand\&{&}

\newcommand{\eq}[1]{~{\rm (\ref{eq:#1})}}

\newcommand{\gDM}{g_{\cal X}}

\newcommand{\GeV}{\,{\rm GeV}}
\newcommand{\TeV}{\,{\rm TeV}}

\newcommand{\Tr}{\,{\rm Tr}}

\def\circa#1{\,\raise.3ex\hbox{$#1$\kern-.75em\lower1ex\hbox{$\sim$}}\,}

\newcommand{\beq}{\begin{equation}}
\newcommand{\eeq}{\end{equation}}

\newcommand{\bea}{\begin{eqnarray}}
\newcommand{\eea}{\end{eqnarray}}
\newcommand{\dif}{d}

\font\tenrsfs=rsfs10 at 12pt
\font\sevenrsfs=rsfs7
\font\fiversfs=rsfs5
\newfam\rsfsfam
\textfont\rsfsfam=\tenrsfs
\scriptfont\rsfsfam=\sevenrsfs
\scriptscriptfont\rsfsfam=\fiversfs
\def\mathscr#1{{\fam\rsfsfam\relax#1}}
\def\Lag{\mathscr{L}}

\def\circa#1{\,\raise.3ex\hbox{$#1$\kern-.75em\lower1ex\hbox{$\sim$}}\,}
\makeatletter

\def\hhref#1{\href{http://arxiv.org/abs/#1}{arXiv:#1}} % in bibliography
\def\baselinestretch{1.09}
\def\hhref#1{\href{http://arxiv.org/abs/#1}{arXiv:#1}}
\usepackage{xstring}
\newcommand{\hhrefq}[1]{\IfSubStr{#1}{:}{\href{http://inspirehep.net/search?ln=en&ln=en&p=#1&of=hb&action_search=Search&sf=&so=d&rm=&rg=25&sc=0}{InSpire:#1}}{\hhref{#1}}}

\def\art{\@ifnextchar[{\eart}{\oart}}
\def\eart[#1]#2#3#4#5#6{{\rm #2}, {\em #3 \bf #4} {\rm (#6) #5} ({\em #1})}
\def\article{\@ifnextchar[{\earticle}{\oarticle}}
\def\oarticle#1#2#3#4#5#6{{\rm #1}, {\ital `#6'}, {\rm #2 #3 (#5) #4}}
\def\earticle[#1]#2#3#4#5#6#7{{\rm #2}, {\ital `#7'}, {\rm #3 #4 (#6) #5}  [\hhrefq{#1}]}
\def\hepart[#1]#2{{\rm #2, \sl#1}}
\def\heparticle[#1]#2#3{#2, {\ital `#3'} [\hhrefq{#1}]}
\newcommand{\doi}[1]{\href{http://dx.doi.org/#1}{[link]}}
\newcommand{\hhrefqq}[1]{\IfBeginWith{#1}{10.}{\href{https://doi.org/#1}{doi:#1}}{\hhrefq{#1}}}

\renewenvironment{thebibliography}[1]
{\begin{multicols}{2}[\section*{\refname}]%
		\@mkboth{\MakeUppercase\refname}{\MakeUppercase\refname}%
		\list{\@biblabel{\@arabic\c@enumiv}}%
		{\settowidth\labelwidth{\@biblabel{#1}}%
			\leftmargin\labelwidth
			\advance\leftmargin\labelsep
			\@openbib@code
			\usecounter{enumiv}%
			\let\p@enumiv\@empty
			\renewcommand\theenumiv{\@arabic\c@enumiv}}%
		\sloppy
		\clubpenalty4000
		\@clubpenalty \clubpenalty
		\widowpenalty4000%
		\sfcode`\.\@m}
	{\renewcommand{\@noitemerr}
		{\@latex@warning{Empty `thebibliography' environment}}%
		\endlist\end{multicols}}

\font\ital=cmu10

\newcommand{\DM}{{\cal X}}
\newcounter{alphaequation}[equation]
\def\thealphaequation{\theequation\hbox to
0.6em{\hfil\alph{alphaequation}\hfil}}
\def\eqnsystem#1{
\def\@eqnnum{{\rm (\thealphaequation)}}
\def\@@eqncr{\let\@tempa\relax \ifcase\@eqcnt \def\@tempa{& & &} \or
  \def\@tempa{& &}\or \def\@tempa{&}\fi\@tempa
  \if@eqnsw\@eqnnum\refstepcounter{alphaequation}\fi
\global\@eqnswtrue\global\@eqcnt=0\cr}
\refstepcounter{equation} \let\@currentlabel\theequation \def\@tempb{#1}
\ifx\@tempb\empty\else\label{#1}\fi
\refstepcounter{alphaequation}
\let\@currentlabel\thealphaequation
\global\@eqnswtrue\global\@eqcnt=0 \tabskip\@centering\let\\=\@eqncr
$$\halign to \displaywidth\bgroup \@eqnsel\hskip\@centering
$\displaystyle\tabskip\z@{##}$&\global\@eqcnt\@ne
\hskip2\arraycolsep\hfil${##}$\hfil& \global\@eqcnt\tw@\hskip2\arraycolsep
$\displaystyle\tabskip\z@{##}$\hfil
\tabskip\@centering&\llap{##}\tabskip\z@\cr}
\def\endeqnsystem{\@@eqncr\egroup$$\global\@ignoretrue} \makeatother

\newcommand{\SU}{\,{\rm SU}}
\newcommand{\SUL}{\SU(2)_L}

\newcommand{\MDM}{M_\DM}
\newcommand{\Fin}{\mathcal{F}}
\newcommand{\Flux}{\mathcal{N}_{\rm flux}}
\newcommand{\Amp}{\mathscr{A}}

\newcommand{\msbar}{\overline{\rm MS}}
\newcommand{\OS}{\rm OS}

\begin{document}
\begin{center}
IFT-UAM/CSIC-26-68
\end{center}
\begingroup
\begin{center}
{\LARGE \bf \color{rossos}
Weak corrections to Minimal\\[1ex] Dark Matter annihilations}\\
\bigskip\bigskip
{\large\bf Dario Buttazzo$^a$},
{\large\bf Mateusz Duch$^b$},\\[0.5ex]
{\large\bf Pier Paolo Giardino$^{c}$},
{\large\bf Alessandro Strumia$^b$}
\\[2ex]
{\it $^a$ INFN Sezione di Pisa, Italia}\\
{\it $^b$ Dipartimento di Fisica dell'Universit{\`a} di Pisa, Italia}\\
{\it $^{c}$  Departamento de F\'isica Te\'orica and Instituto de F\'isica Te\'orica, \\
    Universidad Aut\'onoma de Madrid, Spain}\\

%{\it $^c$ Karlsruhe Institute of Technology, Germany}

\bigskip\bigskip

{\large\bf\color{blus} Abstract}
\begin{quote}\large
We compute the one-loop weak corrections to the annihilation cross sections of fermionic Minimal Dark Matter multiplets.
Infrared divergences cancel in the dominant $s$-wave combination relevant for the thermal relic abundance.
Instead, infrared-enhanced corrections affect velocity-suppressed rates, through a Sudakov/Sommerfeld interplay. The corrections grow with the multiplet size and are at the $5\%$ level in the most motivated cases: the Higgsino-like doublet, the wino-like triplet, the stable quintuplet.
\end{quote}
\thispagestyle{empty}
\end{center}
\tableofcontents
\endgroup
\setcounter{footnote}{0}
%\newpage

\section{Introduction}
Adding a fermionic electroweak $n$-plet to the Standard Model leads to a simple and predictive Minimal framework for Dark Matter (MDM), in which the only free parameter is the dark matter mass $\MDM$~\cite{hep-ph/0512090,2406.01705}.
The renormalizable Lagrangian of the theory is
\beq \label{eq:MDMlagrangian}
\Lag = \Lag_{\rm SM} +c
%\left\{\begin{array}{ll}
 \bar{{\cal X}} (i D\hspace{-1.4ex}/\hspace{0.5ex}+\MDM) {\cal X}
\eeq
where $c=1/2$ when DM is a Majorana fermion with hyper-charge $Y=0$
%(this is possible for odd $n =3, 5,\ldots$)
and $c=1$ when DM is a complex Dirac fermion with $Y\neq 0$.
Multiplets with $Y\neq 0$ are excluded by too large direct detection  tree-level effects, but
the exclusion can be avoided at the price of some non minimality~\cite{2205.04486}.
A weak quintuplet is automatically stable on cosmological timescales~\cite{hep-ph/0512090}.
A triplet with hypercharge $Y=0$ or a doublet with $Y=1/2$ can be made stable through the imposition of a $\mathbb{Z}_2$ symmetry; these cases are  motivated in supersymmetric constructions as the ``wino'' and ``higgsino'' (see e.g.~\cite{hep-th/0405159}).

For each DM multiplet choice,
the value of its mass $\MDM$ is fixed by requiring that thermal freeze-out in standard cosmology reproduces the observed relic abundance, $\Omega_{\rm DM} h^2 = 0.120 \pm 0.001$~\cite{2406.01705}.
Achieving a more accurate prediction for the MDM mass is  important because
 indirect detection signals exhibit a strong dependence on $\MDM$~\cite{0706.4071,1702.01141,2507.17607},
and the dark matter production cross section at a leptonic collider
is  enhanced if the center-of-mass energy is tuned to resonantly produce DM bound states~\cite{2103.12766}.

\smallskip

One loop QCD corrections to dark matter annihilation processes relevant for the relic abundance have been computed in~\cite{2508.02778}.
As expected, IR-enhanced corrections cancel after summing real and virtual effects.
In this work, we evaluate the one-loop electroweak corrections,
which become increasingly important for large multiplet size $n$,  such as $n=5$ or higher.
We focus on fermionic MDM candidates, as  scalar candidates have one extra free parameter,
the scalar quartic coupling, that gets renormalised by electroweak corrections.

Two special kinds of electroweak corrections have been computed in previous works, as each can be large.
First, electroweak Sommerfeld effects~\cite{hep-ph/0610249,0706.4071,2305.01680} (and the related
formation of bound states~\cite{1702.01141,2107.09688})
give ${\cal O}(1)$ corrections to multiplets with $n \geq 3$.
Second, IR-enhanced
Sudakov corrections sizeably affect all multiplets with mass $\MDM \gg M_{W,Z}$
when computing cross sections affected by infrared (IR) enhancements~\cite{1009.0224,1104.2996,1107.4453,1111.2916,1202.0692,1305.6391,1409.7392,1409.8294,1510.02470,1712.07656,1805.07367,1903.08702,2211.14341,2309.11562}.
The computation of the DM relic density involves cross sections summed over all DM components,
so that Sudakov corrections cancel in the unbroken $\SUL$ limit.
We investigate whether this cancellation can be spoiled by Sommerfeld effects.

In section~\ref{sec:SU2s} we compute electroweak corrections to the dominant $s$-wave cross sections in the $\SU(2)_L$-invariant limit, finding that they are IR finite.
In section~\ref{sec:components} we decompose such corrections in components, as needed in some cases for precision computations.
In section~\ref{sec:IR} we explain the previous result and show that velocity-suppressed cross sections are affected by
a Sudakov/Sommerfeld interplay, that we compute.
In section~\ref{sec:235} we specialise our results to the most motivated multiplets, and solve Boltzmann equations
to get predictions for their masses.
Conclusions are given in section~\ref{sec:concl}.

\section{Electroweak corrections in the SU(2)$_L$-invariant limit}\label{sec:SU2s}
At leading tree-level order a pair of DM $n$-plet components, $\DM_i\bar \DM_j$, annihilates into three classes of final states:
two $\SUL$ gauge bosons $WW$, a pair of left-chiral fermion doublets $f \bar f$, or a Higgs-doublet pair $H H^\dagger $.
The annihilation cross section relevant for the thermal relic abundance are summed
over final states including their weak isospin.
However, even in the unbroken phase of the theory (thermal effects restore  $\SUL$ at temperature $T \gtrsim 155\GeV$),
the effective coannihilation rate is not a flat average over initial isospin states.
It contains Sommerfeld factors that depend only on the total isospin $I$ of the DM pair.

In this section we focus on the $s$-wave effects, that dominate in the limit $v\to 0$, giving
constant cross-sections annihilations times relative DM velocity $v$.
Annihilations into two $\SUL$ vectors $WW$ proceed as $I=1$ and $I=5$
(while $I=3$ receives a sub-dominant $v^2$ contribution, and $I>5$ is not allowed by the final state);
annihilations into fermions and Higgs proceed as $I=3$.
The relevant cross section is averaged
over DM {\em and} $\overline{\rm DM}$ components,
such that the observed dark matter abundance is reproduced for $\sigma v \approx 1/(23\TeV)^2$~\cite{2406.01705}.
The tree level $s$-wave cross sections are~\cite{hep-ph/0512090,0706.4071,1702.01141}
\beq\begin{split}
\sigma v^{WW}_{I=1} = \frac{\pi \alpha_2^2 (n^2 - 1)^2}{24 \MDM^2 \gDM},\qquad
&\sigma v^{WW}_{I=5} =\frac{\pi \alpha_2^2 (n^2 - 4)(n^2 - 1)}{12 \MDM^2 \gDM},\cr
\sigma v^{f\bar{f}}_{I=3} = n_L \frac{\pi \alpha_2^2 (n^2 - 1)}{8 \MDM^2 \gDM},\qquad
 & \sigma v^{HH^\dagger }_{I=3} = \frac{\pi \alpha_2^2 (n^2 - 1)}{16 \MDM^2 \gDM},
\label{eq:SigmaTree}
\end{split}\eeq
where $n_L=12$ is the number of $\SUL$ fermionic doublets, quarks and leptons.
The factor $\gDM$ is the number of degrees of freedom in the DM multiplet:
$\gDM=2n~(4n)$ when DM is a Majorana (Dirac) fermion.
For brevity, we here omitted the extra tree level effect present when $Y\neq 0$ and
involving the hyper-charge coupling $g_Y$~\cite{hep-ph/0512090,0706.4071,1702.01141}.
% is here omitted, as we will neglect it in the loop correction.
%Indeed it is relevant only for Dirac multiplets with $Y\neq 0$ and is not enhanced by $n$.
%Cross-sections in other possible channels $\sigma v^{WW}_{I=3}$, $\sigma v^{f\bar f}_{I=1}$ and $\sigma v^{HH^\dagger }_{I=1}$  vanish in the $s$-wave limit.
Including Sommerfeld corrections $S_I$, the total DM annihilation cross section at tree level is
\beq \sigma v|_{\rm tree} = S_1\,  \sigma v_1^{WW} +
S_3 [ \sigma v_3^{f\bar{f}}  +\sigma v^{HH^\dagger }_3]+
S_5 \, \sigma v_5^{WW} .\eeq
Extra processes involving DM bound states are relevant for $n>3$ and can be reduced to
one extra term in the annihilation rate~\cite{1702.01141}; these effects have been computed in~\cite{2107.09688,2205.04486} for all multiplets, we leave the issue of their EW corrections to future work.
The tree-level weak potential is $V =  \alpha_{\rm eff}  e^{-M_V r}/r$ where
$\alpha_{\rm eff}  =  \alpha_2 (\bar\mu)(I^2+1-2n^2)/8$ and $M_V$ is the vector thermal mass~\cite{1702.01141}.
Setting the renormalisation scale at $\bar \mu\sim 1/r$ as done in~\cite{1702.01141}
roughly captures higher order SM corrections to the weak potentials and thereby to $S_I$~\cite{2009.00640,2305.01680}.

%Eq.~(28) of~\cite{1702.01141} provides an analytic approximation for the Abelian Sommerfeld factors;
%neglecting $M_V$ it simplifies to
%\beq S_I = \frac{2\pi \alpha_{\rm eff} /v}{1-e^{-2\pi \alpha_{\rm eff} /v}}.\eeq

Including one-loop corrections, we  write the Sommerfeld-corrected $s$-wave annihilation cross section as
\beq\label{eq:sigmaLNO}
\begin{array}{rcl} \sigma v|_{\rm NLO} &=&  S_1 \,\sigma v_1^{WW} (1+\delta_1^{WW})
+ S_5\, \sigma v_5^{WW}  (1+\delta_5^{WW})
+ S_7\, \delta\sigma v_7^{WWW}+\\
&& \displaystyle S_3\, \left[
\sigma v_3^{f\bar f}  (1+\delta_3^{f\bar f} ) +
\sigma v_3^{H H^\dagger}  (1+ \delta_3^{HH^\dagger} )+
\delta \sigma v_3^{WWW}\right].
\end{array}\eeq
Eq.\eq{sigmaLNO} involves additional effects $\delta( \sigma v)$ with total isospin $I=3,7$,
which become allowed when DM annihilates into three weak vectors.
Moreover, the rates already present at tree level receive both real corrections,
from vector emission, and virtual corrections, arising from the interference between tree-level and one-loop amplitudes.
We parameterize these contributions as
\beq
\delta^\Fin \equiv \frac{\sigma^\Fin_{\rm NLO}-\sigma^\Fin_{\rm tree}}{\sigma^\Fin_{\rm tree}} =
\delta_{\rm real}^\Fin + \delta_{\rm vir}^\Fin.
%\quad\quad
%\delta_V^\Fin \equiv \frac{\Delta\sigma^\Fin_V}{\sigma_{LO}},\quad\quad \delta_R^\Fin \equiv \frac{\Delta\sigma^\mathcal{R}_V}{\sigma_{LO}},
\label{eq:deltadef}
\eeq
%\beq \delta_{\rm QCD}^{q\bar q} =
%.\eeq
% The general expression is delta = (3/4) q^2 alpha/π for fermions and 4 times larger for scalars
All the $\delta$ corrections turn out to be separately IR-convergent, for the reason explained later in section~\ref{sec:IR}.
The extra channels opened by purely real corrections do not contain any IR divergence:
\beq \label{eq:WWW}
\begin{split}
 \delta\sigma v^{WWW}_{3} &= \frac{\alpha_2^3 (n^2 - 1) \left[ n^2 (3n^2 - 14)(\pi^2 - 9) + 133\pi^2 - 1297 \right]}{120 \MDM^2 \gDM},\\
 \delta\sigma v^{WWW}_{7} &= \frac{\alpha_2^3 (n^2 - 1)(n^2 - 4)(n^2 - 9) (\pi^2 - 9)}{60 \MDM^2 \gDM}.
\end{split}
\eeq
Processes already present at tree level are affected by real and virtual corrections.
After summing them we find that IR-enhanced effects cancel separately in each $\delta$, leaving
{\small\begin{eqnsystem}{sys:delta}
\delta^{W\! W}_{1} &= &\frac{\alpha_2}{\pi}\left[\frac{1}{2}\beta_2^{\rm SM}{\cal L}+\frac{\pi^2 (n^2 - 1)}{8 \beta} + \frac{ 3320 - {64} n_L - 165 \pi^2 + 9 n^2 (\pi^2 - 20)}{144} \right], \\
\delta^{W\! W}_{5} &= &\frac{\alpha_2}{\pi}\left[\frac{1}{2}\beta_2^{\rm SM}{\cal L}+\frac{\pi^2(n^2 - 13)}{8 \beta} + \frac{5480 - {64} n_L - 309 \pi^2 + 9 n^2 (\pi^2 - 20)}{144} \right] ,\\
\delta^{f\!\bar{f}}_{3} &= &\delta_3+   \frac{3\alpha_3}{4\pi} + \frac{\alpha_Y}{16\pi}+\frac{\alpha_t}{48\pi}, \label{eq:delta3ff}\\
\delta^{H\!H^\dagger }_{3} &= &\delta_{3} +\frac{11\alpha_2}{16\pi}+ \frac{3 \alpha_Y}{4\pi}-\frac{11 \alpha_t}{4\pi},
%\frac{\alpha_2}{\pi}\left[\frac{1}{2}\beta_2^{\rm SM}\ell+\frac{\pi^2 (n^2 - 5)}{8 \beta}  -\frac{4c n (n^2 - 1)}{27} - \frac{36 n^2 + {10} n_L + 18 \pi^2 - 605 - 48 \ln2}{36}\right] .
\end{eqnsystem}}
where
\beq
\delta_3 =\frac{\alpha_2}{\pi}\left[\frac{1}{2}\beta_2^{\rm SM}{\cal L}+\frac{\pi^2 (n^2 - 5)}{8 \beta}   - \frac{144 n^2 + 40 n_L + 72 \pi^2 - 2321 - 192 \ln2}{144} -\frac{\gDM (n^2 - 1)}{27} \right]\eeq
and $\alpha_t=y_t^2/4\pi$ is the top Yukawa coupling.
The QCD correction in eq.\eq{delta3ff} affects quarks only and was computed in~\cite{2508.02778}.
We added the similar correction arising from the hyper-charges of SM fermions and scalar doublets.
Extra hyper-charge corrections arise if DM has $Y\neq 0$, as discussed later in section~\ref{sec:2}.
The first term in each correction factor $\delta$ is proportional to the weak gauge beta function $\beta_2^{\rm SM}$
%of eq.~(\ref{eq:couplingct})
times ${\cal L}\equiv \ln(\bar\mu^2/4\MDM^2)$.
It cancels the dependence of the tree level cross section $\sigma v|_{\rm tree} \propto \alpha_2^2(\bar\mu)$
on the renormalization scale $\bar\mu$.
No analogous factor is present  for the DM mass $\MDM$ because we renormalize it as pole mass.
The term proportional to $\gDM$ in $\delta_3$ arises from DM corrections to the vector propagator.
The terms enhanced by $1/\beta$, where $\beta= \sqrt{s/4\MDM^2-1} = v/2 \ll 1$ is the DM velocity in the center of mass frame,
are leading Sommerfeld effects.
Such effects %agree with~\cite{1702.01141} and 
must be omitted from the $\delta$ and resummed into Sommerfeld factors $S_I$.
Extra details of the computation are provided in the appendix~\ref{sec:appA}.

\begin{figure}[t]
$$\includegraphics[width=0.8\textwidth]{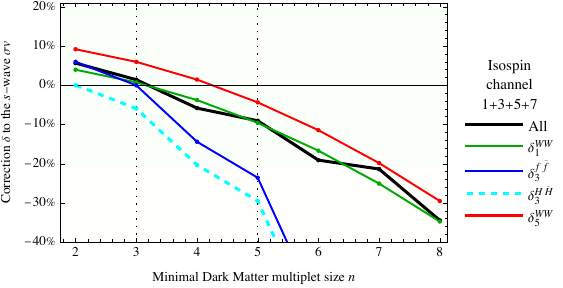}$$
\begin{center}
\caption{\em One loop EW, QCD and top Yukawa corrections to the annihilation cross sections of a Minimal DM $n$-plet of $\SUL$.
The $\delta$ corrections
are computed with respect to the tree level cross section $\sigma \sim \alpha_2^2/\MDM^2$
written in terms of the pole mass $\MDM$ and of $\alpha_2$ renormalized at  $\mub=2\MDM$.
The correction to the total cross section includes the extra $I=3,7$ channels
and approximate Sommerfeld factors.
The thermal mass roughly scales as $\MDM\propto\sqrt{\sigma}$.
\label{fig:EWcorrectionMDM}}
\end{center}
\end{figure}

Fig.\fig{EWcorrectionMDM} illustrates the numerical size of the various $\delta$ corrections as function of the multiplet size
$n$.
As expected weak corrections grow with $n$, and a quintuplet is in the perturbative regime.
The figure also shows an approximate total $\delta$, that corrects the total cross section,
having combined all isospin channels (including eq.\eq{WWW})
by
inserting approximated values for the Sommerfeld factors $S_I$ at the decoupling temperature $T \sim \MDM/25$.
The relic DM abundance roughly scales as $\Omega_{\rm DM}\propto 1/\sigma$.
So the DM mass that matches the observed $\Omega_{\rm DM}$ scales as $\MDM\propto\sqrt{\sigma}$:
a $10\%$ increase in the cross section leads to a $5\%$ increase in $\MDM$.
Precise predictions for $n=\{2,3,5\}$ will be presented in section~\ref{sec:235},
running the relevant Boltzmann equations for DM freeze-out.

\section{Electroweak corrections to DM components}\label{sec:components}
Here we apply the previous results, derived in the $\SUL$-symmetric limit, to the regime in which $\SUL$ is broken and argue that 
the symmetry breaking effects neglected in the calculation of annihilation cross-section in the previous section give only sub-leading corrections.
This extension is required for the computation of cosmological dark-matter annihilation processes below the electroweak phase-transition temperature.
In this regime, electric charge, rather than isospin, provides the relevant conserved quantum number.
We found that the dominant $s$-wave annihilation rates receive finite EW and QCD corrections,
thereby negligibly affected by small vector masses $M_{W,Z}\ll \MDM$.
Nevertheless, the component computation gets more involved because 
the weak-interaction potentials governing the Sommerfeld effect acquire a matrix structure.
Then, to properly account for the relative phases among the Sommerfeld-corrected states,
one needs to go beyond annihilation cross sections and compute
the imaginary part $\Gamma$ of the propagators of the two-body DM states.
In view of the optical theorem, its diagonal components reproduce the annihilation cross sections.
Its off-diagonal components are needed to take into account interferences
among states mixed by Sommerfeld corrections~\cite{hep-ph/0610249,0706.4071}.

We thereby describe the $\SU(2)_L$ algebra that allows to first decompose amplitude
into total isospin channels (as done in section~\ref{sec:SU2s}), and next to project the result over components.
The matrix element for an annihilation of a pair of DM states into a final state $\Fin$ is\beq
\Amp^{\Fin}_{m_1m_2}= \langle\Fin|\Amp|  m_1, m_2 \rangle,
\eeq
where $m_1,m_2\in\{-(n-1)/2,\ldots,(n-1)/2\}$ are the $T_3$  eigenvalues that label the components of a DM multiplet with dimension $n$.\footnote{For a Dirac multiplet, the annihilating pair is $\chi_{m_1}\bar\chi_{m_2}$, where $\bar\chi_{m_2}$ denotes the antiparticle of $\chi_{-m_2}$.}
% where $\alpha$ collectively denotes the final-state $\SUL$ indices.
This amplitude can decomposed into irreducible representations of dimension
$I \in\{1,3,\ldots,2n-1\}$ labeled by an index $M\in\{-(I-1)/2,\ldots,(I-1)/2\}$.
The decomposition is effected by the Clebsch-Gordan tensors
\beq {\mathcal C^{IM}_{m_1m_2}}=\langle m_1,  m_2 | I,M\rangle\qquad\hbox{so that}
\qquad
\Amp^{\Fin}_{IM} = \langle \Fin | \Amp | I,M \rangle = \sum_{m_1,m_2}\mathcal  \Amp^{\Fin}_{m_1m_2} \mathcal C^{IM}_{m_1m_2}.
\label{eq:amp_isospin}
\eeq
The cross-section for the annihilation into a final state $\Fin$ is
\beq
\sigma v_{IM}^\Fin %=  \Flux \int d\Phi^{\Fin} \langle I,M | \Amp^\dagger |F\rangle\langle F| \Amp |I,M \rangle
=  \Flux \int d\Phi^{\Fin} |\Amp^{\Fin}_{IM}|^2,
\label{eq:sigmav_IM}
\eeq
where $\Flux$ denotes the flux factor.
%The standard averaging over initial state spins and summation over final state degrees of freedom, including symmetry factors for identical particles in the final state, is implicitly assumed. \AS{Should be defined better}
The more general matrix $\Gamma$ needed to include Sommerfeld corrections is
%the relevant object is not only the cross-section given by short-distance physics. One needs the full hard annihilation matrix $\Gamma$ that is proportional to the absorptive part of the forward scattering amplitude $\Gamma\sim -i(\Amp_{\rm fwd}-\Amp^\dagger_{\rm fwd})$ or equivalently () is given by
\beq
 \Gamma^\Fin =   \Flux \int  d\Phi^{\Fin}  \Amp^\dagger | \Fin \rangle\langle \Fin | \Amp.
\eeq
These $\Gamma$ matrices are used to compute the annihilation cross-section averaged over the multiplet
in the presence of Sommerfeld factors described by a matrix $A$ of wave-functions at the origin as~\cite{0706.4071}
$
 \sigma v = \Tr A \Gamma A^\dagger$.
Small $\SUL$-breaking effects can be neglected in $\Gamma$, but significantly affect wave-functions at low energies,
comparable to the mass splittings among the DM components. 
We approximated the  flux factor as common to all states, as mass splittings are small with respect to $\MDM$.

The electric charge of the initial pair $Q=M$ is conserved, therefore $\langle I',M'|\Gamma| I,M \rangle$ can be non-vanishing only when $M=M'$.
The projector $ | \Fin \rangle\langle \Fin | $ becomes $\SUL$ invariant after summing over all the $\SUL$ components of the final state.
Then, considering the amplitudes in the unbroken theory, the matrix $\Gamma^\Fin$ is diagonal also in $I$ and has a simple form\footnote{A different projector arises when considering a specific final state, as relevant in indirect detection rates, such that, in general,  its annihilation matrix contains interference terms between different
initial isospin channels.}
\beq
\Gamma^\Fin = \sum_{I} \Gamma^\Fin_{I} \, \Pi^I,
\label{eq:Gamma}
\eeq
where we introduced the projector on the subspace with isospin $I$: $\Pi^I=\sum_M |I,M\rangle\langle I,M|$ and
$\Gamma^\Fin_I=\sigma v_{IM}=\sigma v_{I}/I$ is the annihilation cross-section for each state in subspace of dimension $I$. The non-diagonal terms that connect initial states with $I'\neq I$ arise only due to $\SU(2)_L$-breaking effects, so they are suppressed by $M^2_{W,Z,t}/\MDM^2$.
In the component basis, we get\footnote{This `spherical' basis is convenient because it corresponds to states with given electric charge.
The amplitudes are computed in terms of the DM $\SUL$ generators $T^a_{ij}$
with  $a\in\{1,2,3\}$, thereby using  a `cartesian' component coordinates $i,j\in\{1,\ldots,n\}$.
Such components are similarly decomposed using the cartesian version of Clebsch-Gordan tensors
$\mathcal{P}^{I,A}=\langle i,j|I,A\rangle$, where $A$ is a generalized tensor index (null for the singlet,
$a$ for the triplet $I=3$, $ab$ for the quintuplet $I=5$, etc), given by
\begin{flalign}
%\;\;& \resizebox{0.9\textwidth}{!}{$\displaystyle
\begin{alignedat}{3}
& \mathcal{P}^{1}_{ij} = \frac{\delta_{ij}}{\sqrt{D_R}}, \\
& \mathcal{P}^{3,a}_{ij} = \frac{T^{a}_{ij}}{\sqrt{N_1}}, \quad &&
  N_1 = T_R = \frac{n(n^2-1)}{12}, \\
& \mathcal{P}^{5,ab}_{ij} = \frac{S^{ab}_{ij}}{\sqrt{N_2}}, \quad &&
  N_2 = \frac{2(4C_R - 3)}{5} T_R, \quad &&
  S^{ab}_{ij} = \{T^a,T^b\}_{ij} - \frac{2C_R}{3}\delta^{ab}\delta_{ij}, \\
& \mathcal{P}^{7,abc}_{ij} = \frac{C^{abc}_{ij}}{\sqrt{N_3}}, \quad &&
  N_3 = \frac{3(4C_R - 3)(C_R - 2)}{70} T_R, \quad &&
  C^{abc}_{ij} = \left[ (T^{\{a} T^b T^{c\}})_{ij} - \frac{3C_R - 1}{15} ( \delta^{ab} T^c_{ij} + \dots )\right]
\end{alignedat}
%$} &
\end{flalign}
where $T_R$ is the Dynkin index and $C_R =  \sfrac{(n^2-1)}{4} $ is the quadratic Casimir of the weak $n$-plet.}
\beq
\langle m'_1 ,m'_2|\Gamma^\Fin| m_1 , m_2\rangle = \sum_{I}\Gamma^\Fin_{I}
\mathcal C^{IM}_{m'_1m'_2} \mathcal C^{IM}_{m_1m_2}.
\label{eq:JM2ellm}
\eeq
The  rates $  \Gamma =  \sum_\Fin \Gamma^\Fin$  in each isospin channel are related to
the  cross sections of section~\ref{sec:SU2s} as
\beq\begin{split}
\Gamma_{I=1} &= \sigma v_{I=1}^{WW} (1+\delta_1^{WW}),
\qquad
\Gamma_{I=5} = \frac15  \sigma v_{I=5}^{WW} (1+\delta_5^{WW}),
\qquad
\Gamma_{I=7} = \frac{1}{7}\,\delta\sigma v^{WWW},
\\
\Gamma_{I=3} &= \frac13 \bigg[
\sigma v_3^{f\bar f}  (1+\delta_3^{f\bar f})+
\sigma v_3^{H H^\dagger}  (1+ \delta_3^{HH^\dagger})+
\delta \sigma v_3^{WWW}\bigg].
\end{split}\eeq
For a Majorana DM multiplet $M=Q$ and it is convenient to go to the `spin basis' of $s$-wave states defined for $m_1<m_2$ by
\beq
 \ket{m_1 ,m_2}_{S}  =  \frac{ \ket{m_1,m_2}  +    (-1)^S\,
    \ket{m_2,m_1}}{\sqrt2}.
%    \otimes \ket{L,S},
  \label{eq:unordered_channel_distinct}
\eeq
For $m_1=m_2$, only the symmetric combination exists, and it equals the component basis state,
$\ket{m_1,m_2}_{S=1}=  \ket{m ,m}$.
%   \otimes\ket{L,S}.
%  \label{eq:unordered_channel_identical}
%\eeq
So the total isospin states are expressed in the spin basis as
\begin{equation}
  \ket{I,M=Q}
  =
  \sum_{m_1 , m_2}
  \tilde{\mathcal C}^{IQ}_{{m_1,m_2}}\,
  \ket{m_1,m_2}_S,\qquad
      \tilde{\mathcal C}^{IQ}_{{m_1m_2}}
  =
  \frac{1-(-1)^{S+n+(I-1)/2}}{\sqrt{2(1+\delta_{m_1m_2})}}\,
  {\mathcal C}^{IM}_{m_1,m_2}.
  \label{eq:IM_unordered_CG}
\end{equation}
%where the form of unitary transformation $U^{I,Q}_{\ket{m_1,m_2}_S}$ can be read from Eqs.~(\ref{eq:unordered_channel_distinct}), (\ref{eq:unordered_channel_identical})
%\begin{equation}
%
%  \label{eq:U_CG_charge_basis}
%\end{equation}
having used $\mathcal C^{IM}_{ m_2, m_1}=-(-1)^{n+(I-1)/2} \mathcal C^{IM}_{ m_1, m_2}$.
This selects even $S+n+(I-1)/2$ as required by the Fermi-Dirac statistics.
The annihilation matrix in the spin basis reads as eq.\eq{JM2ellm} with ${\cal C}\to\tilde{\cal C}$.
%\AS{old:
%\beq
%\prescript{}{S}{\bra{m'_1m'_2}} \Gamma \ket{m_1m_2}_S = \sum_{I} \Gamma_I   \tilde{\mathcal C}^{I,Q}_{\ket{m'_1m'_2}_S}\,
%  \tilde{\mathcal C}^{I,Q}_{\ket{m_1m_2}_S}.
%\eeq}

%Additional mass splittings corrections in the coannihilation rate can be added removing the multiplet average factor $1/n^2$ in the cross-sections and inserting the thermal density matrix in the trace above.

We give explicit expressions in the triplet and quintuplet case, using
the same compact notation as in~\cite{0706.4071}, where the indices $m_1, m_2$ labeling the matrices denote the diagonal elements $\prescript{}{S}{\bra{m_1m_2}}\Gamma\ket{m_1m_2}_S$,
while the off-diagonal entries are left implicit.\footnote{Some literature \cite{hep-ph/0610249,0706.4071,Beneke:2014gja} uses $\hat\Gamma=\Gamma/2$.
For $m_1\neq m_2$ it satisfies $\bra{m_1,m_2}\Gamma\ket{ m_1,m_2} = \bra{m_2,m_1}\Gamma\ket{ m_2 ,m_1} = \sum_{S=0,1}\prescript{}{S}{\bra{m_1m_2}}\hat\Gamma\ket{m_1,m_2}_S$. For $m_1 = m_2 = m$ it satisfies $\bra{m,m}\Gamma\ket{m;m}=2\prescript{}{S=0}{\bra{m,m}}\hat\Gamma \ket{m,m}_{S=0}$.
%This is the origin of the extra factor of~$2$ in the annihilation rate for the same DM components $m_1=m_2$ that arises when one identifies $\hat\Gamma$ elements in the spin basis with elements of $\Gamma$ in the component basis~\cite{hep-ph/0610249,0706.4071,Beneke:2014gja}.
}
For the $n=3$ triplet we find
\begin{eqnsystem}{sys:Gamma3}
\Gamma _{Q=0}^{S=0} &=& \bordermatrix{&{\scriptstyle +}&{\scriptstyle 0}\cr
{\scriptstyle -}& \sfrac{2 \Gamma _1}{3}+\sfrac{\Gamma _5}{3} & \sfrac{\sqrt{2} \Gamma _1}{3}-\sfrac{\sqrt{2} \Gamma _5}{3} \cr
{\scriptstyle 0}& \sfrac{\sqrt{2} \Gamma _1}{3}-\sfrac{\sqrt{2} \Gamma _5}{3} & \sfrac{\Gamma _1}{3}+\sfrac{2 \Gamma _5}{3} }
,\\
\Gamma _{Q=0}^{S=1} &=&  \Gamma _{Q=1}^{S=1}  = \Gamma_3
,\\
\Gamma _{Q=1}^{S=0} &=& \Gamma _{Q=2}^{S=0} = \Gamma_5
\end{eqnsystem}
For the $n=5$ quintuplet we find
\begin{eqnsystem}{sys:Gamma5}
 \Gamma _{Q=0}^{S=0} &=& \bordermatrix{&{\scriptstyle ++}&{\scriptstyle +}&{\scriptstyle 0}\cr
{\scriptstyle --} & \sfrac{2 \Gamma _1}{5}+\sfrac{4 \Gamma _5}{7} & \sfrac{2 \Gamma _1}{5}-\sfrac{2 \Gamma _5}{7} & \sfrac{\sqrt{2} \Gamma _1}{5}-\sfrac{2 \sqrt{2} \Gamma _5}{7} \cr
{\scriptstyle -}& \sfrac{2 \Gamma _1}{5}-\sfrac{2 \Gamma _5}{7} & \sfrac{2 \Gamma _1}{5}+\sfrac{\Gamma _5}{7} & \sfrac{\sqrt{2} \Gamma _1}{5}+\sfrac{\sqrt{2} \Gamma _5}{7} \cr
{\scriptstyle 0}& \sfrac{\sqrt{2} \Gamma _1}{5}-\sfrac{2 \sqrt{2} \Gamma _5}{7} & \sfrac{\sqrt{2} \Gamma _1}{5}+\sfrac{\sqrt{2} \Gamma _5}{7} & \sfrac{\Gamma _1}{5}+\sfrac{2 \Gamma _5}{7}}
,\\
\Gamma _{Q=0}^{S=1} &=& \bordermatrix{&{\scriptstyle ++}&{\scriptstyle +}\cr
{\scriptstyle --}& \sfrac{4 \Gamma _3}{5}+\sfrac{\Gamma _7}{5} & \sfrac{2 \Gamma _3}{5}-\sfrac{2 \Gamma _7}{5} \cr
 {\scriptstyle -}&\sfrac{2 \Gamma _3}{5}-\sfrac{2 \Gamma _7}{5} & \sfrac{\Gamma _3}{5}+\sfrac{4 \Gamma _7}{5} }
,\\
\Gamma _{Q=1}^{S=0} &=&  \bordermatrix{& {\scriptstyle ++} & {\scriptstyle +}\cr
{\scriptstyle -}& \sfrac{6 \Gamma _5}{7} & \sfrac{\sqrt{6} \Gamma _5}{7} \cr
{\scriptstyle 0}& \sfrac{\sqrt{6} \Gamma _5}{7} & \sfrac{\Gamma _5}{7} }
,\\
\Gamma _{Q=1}^{S=1} &=& \bordermatrix{& {\scriptstyle ++} & {\scriptstyle +}\cr
{\scriptstyle -}&  \sfrac{2 \Gamma _3}{5}+\sfrac{3 \Gamma _7}{5} & \sfrac{\sqrt{6} \Gamma _3}{5}-\sfrac{\sqrt{6} \Gamma _7}{5} \cr
{\scriptstyle 0}& \sfrac{\sqrt{6} \Gamma _3}{5}-\sfrac{\sqrt{6} \Gamma _7}{5} & \sfrac{3 \Gamma _3}{5}+\sfrac{2 \Gamma _7}{5} }
,\\
\Gamma _{Q=2}^{S=0} &=& \bordermatrix{&{\scriptstyle ++}&{\scriptstyle +}\cr
{\scriptstyle 0}& \sfrac{4 \Gamma _5}{7} & -2 \sqrt{3} \Gamma _5/7\cr
{\scriptstyle +}&-2 \sqrt{3} \Gamma _5/7 & \sfrac{3 \Gamma _5}{7} },\\
\Gamma _{Q=2}^{S=1} &=& \Gamma _{Q=3}^{S=1} = \Gamma_7
,\\
\Gamma _{Q=3}^{S=0} &=&  \Gamma _{Q=4}^{S=0} =0.
\end{eqnsystem}
In conclusion, we derived explicit expressions for the one-loop corrections to the annihilation matrices
used in Sommerfeld computations.

\section{IR-enhanced corrections}\label{sec:IR}
In this section we discuss a general issue in quantum field theory:
Sudakov and Sommerfeld corrections can be separately large, needing
different resummations that might clash with each other, leaving enhanced corrections.

Two kinds of large infrared logarithms can arise: soft and collinear.
When both are present, their overlap leads to double Sudakov logarithms, which exponentiate.
Otherwise, single Sudakov logarithms arise.
As a result of large Sudakov logarithms, TeV-scale electroweak multiplets receive order-unity electroweak corrections,
 $\alpha_2\ln^2(\MDM/M_W)/4\pi \sim 1$~\cite{1009.0224,1305.6391}.
However,  the KLN theorem~\cite{Kinoshita:1962ur,Lee:1964is} guarantees that infrared singularities
and their associated Sudakov corrections cancel after combining real with virtual corrections
to sufficiently inclusive observables, that involve summing over degenerate final and initial states.
In our case, this amounts to combining the one-loop corrections to the \(2\to2\) process with the tree-level \(2\to3\) amplitudes involving the emission of an additional soft gauge boson. 
Physically, the virtual corrections subtract the probability for unresolved soft radiation from the exclusive hard process with no real emission.
On the other hand, fixed initial isospin states retain Bloch-Nordsieck violating logarithms, 
while an average over the full initial degenerate multiplet cancels the leading logarithmic correction \cite{Ciafaloni:2000df,hep-ph/0004071}. 

The applicability of the KLN argument to  DM relic-abundance calculations is subtle.
%Although one performs an inclusive sum over final states, soft bosons are not included in the sum over initial states. \AS{why not?}

The KLN cancellation is spoiled in the thermal co-annihilation rate by non-zero mass splittings that arise 
at low temperatures where $\SUL$ is broken \cite{1305.6391}. 
Power suppressed logarithms $\sim \alpha_2 M^2_W/\MDM^2 $ are also not guaranteed to cancel \cite{2202.00934}. 
However, these effects are small in view of the small mass splittings predicted by multi-TeV minimal DM theories.

Here we focus on a potentially larger effect, that is present already in the unbroken $\SUL$ high temperature environment. 
Naively, multiplying annihilation rates for each initial isospin by the channel-dependent Sommerfeld factors spoils the cancellation of logarithmic corrections. 
We will show that such enhancements did not affect our previous computation of the leading $s$-wave rates because no soft radiation is emitted for $v\to 0$,
while velocity-suppressed terms \cite{1009.0224,1305.6391} are affected.

\smallskip

We analyze the real contributions in section~\ref{sec:real}. Given the structure of the infrared cancellations, this is sufficient to clarify the issue.
In section~\ref{sec:virtual} we then extend the discussion to virtual corrections, summarising the formalism that allows to derive infrared-enhanced effects.
This is applied to velocity-suppressed cross sections in section~\ref{sec:pwave}.

%For example, when two DM particles annihilate at tree level into a pair of $\SUL$ vectors, the allowed total isospin are $I =1$ and $I=5$.
%At higher order one has virtual corrections at one loop (where the same isospin channels arise) plus
%real emission into 3 vectors (that opens the $I=7$ channel and makes $I=3$ no longer forbidden by anti-symmetry).
%So it's not clear if/how IR effects cancel.
%If the KLN cancellation only happens after summing over all $I$, then it can be broken by Sommerfeld.

\subsection{Real corrections}\label{sec:real}
The initial pair ${\cal X}(p_1)\bar{\cal X}(p_2)$ is massive, so no collinear enhancement arises
and infrared singularities arise solely from soft dynamics.
%Collinear radiation is associated with individual legs and does not probe the pair collectively:
%it would see the opposite direction of the two particles.
%But collinear radiation is negligible in the non-relativistic limit.
%Formally, the probability of emitting one extra vector from a particle with momentum $p$ and mass $M$
%is obtained multiplying the squared amplitude by the emission probability given by the splitting function
%\beq d\wp \propto P(x) dx \frac{dk_\perp^2}{k_\perp^2 + x^2 M^2}\eeq
%showing that no divergence arises in the collinear limit $k_\perp \to 0$.
%The variable $x$ is the fraction of the parent particle’s longitudinal momentum retained after emission.
In the soft limit, the emission amplitude of a soft weak vector boson $W^a(k)$ with small momentum $k_\mu = \omega (1, n_x, n_y, n_z)$
and polarization $\varepsilon^\mu$
factorizes into the hard amplitude times the eikonal factor
%$J_\mu^a = \sum_{i=1}^2 T_i^a p_{i\mu}/(p_i\cdot k)$.
\beq
 R_a(k)
  =
  g_2\,\varepsilon_\mu^*
  \left[
  \frac{p_1^\mu}{p_1\cdot k}T_1^a
  +
  \frac{p_2^\mu}{p_2\cdot k}T_2^a
  \right]=
  \frac{g_2}{2}\varepsilon_\mu^*
  \left[J_+^\mu(k)T^a_+ +J_-^\mu(k)T^a_-\right],
  \label{eq:R_Jpm}
  \eeq
where $T^a_{1,2}$ are the isospins of the two DM particles, and 
\beq
J_\pm^\mu(k)
  =
  \frac{p_1^\mu}{p_1\cdot k}
  \pm
  \frac{p_2^\mu}{p_2\cdot k}\,,\quad\quad\quad T^a_\pm=T^a_1 \pm T^a_2\,.
\eeq
The `monopole' part of the eikonal current is proportional to the total isospin generator $T^a_+$. It only rotates states within a fixed initial isospin $I$ representation and therefore does not induce transitions between different Sommerfeld channels. As will shown later, this part does not lead to any soft correction, because its real and virtual contributions cancel. This is consistent with the fact that in the physical gauge transverse emission from the monopole vanishes.

\smallskip

The dipole current $J_-$ couples to  $T^a_-=T_1^a-T_2^a$, which is
a vector under total isospin and connects neighbouring representations, $I\to I\pm2$. It is a conserved current $k_\mu J^\mu_- =0$ related to the physical emission.  In the non-relativistic limit
\beq
p_1^\mu \simeq \MDM(1,0,0,\beta),
\qquad
p_2^\mu \simeq \MDM(1,0,0,-\beta),
\qquad
\beta \ll 1.
\eeq
we can square the amplitude, sum over polarizations and integrate over the phase space to get the emission rate expanded in $\beta$
\begin{equation}
  \calW_-
  =
  \frac{g^2_2}{4}\int d\Phi_k
  \left[-J_-^2(k)\right]
  =
  -\frac{\alpha_2}{4\pi}
\left(\frac{4\pi\bar\mu^2}{s}\right)^{\epsilon}
\frac{1}{\Gamma(1-\epsilon)}
\left[
\frac{4\beta^2}{3\epsilon}
+{\cal O}(\beta^4)
\right].
  \label{eq:Wpm_def}
\end{equation}
where $ d\Phi_k =
  \bar\mu^{2\epsilon}\sfrac{\dif^{d-1}k}{(2\pi)^{d-1}2\omega}$ is the phase space measure in $d$ dimensions.
%The mixed $J_+ J_-$ term vanishes after angular integration, leaving
%The eikonal factors are 
%\MD{$J_+$ is not conserved, so the polarization vectors should not be summed with $g_{\mu\nu}$ in the square. It has zero total effect on the annihilation matrix, so it can be neglected. Separately, for real emission it makes sense only in axial gauge with $n=p_1+p2$, and then there is no radiation, this agrees with the fact that monopole does not radiate.}
We note that IR-enhanced initial state radiation corrections are velocity supressed and do not affect the dominant $s$-wave cross sections \cite{1305.6391}.
% and affect the $p$-wave cross sections.

\subsection{Virtual corrections}\label{sec:virtual}
We next include virtual corrections.
The scattering operator $S$ can be factorized in terms of the hard scattering operator $\SH$ times the
operators $U_I$ and $U_F$ describing the initial and final-state soft dynamics induced by the eikonal coupling:
 emission of soft vectors and rotation of hard charges~\cite{Ciafaloni:2000df,hep-ph/0004071}
\begin{equation}
  S=U_F\,\SH\,U_I .
  \label{eq:S_factorized}
\end{equation}
Using this factorized form, the KLN theorem follows from the unitarity of $U_I$ and $U_F$ which drop out from the sum over complete sets of degenerate initial $i$ and final $f$ states,
  including unresolved soft quanta~\cite{Bloch:1937pw,Kinoshita:1962ur,Lee:1964is}:
\begin{equation}
  \sum_{i,f}  |\langle f|S|i\rangle|^2 =  \Tr_{{\rm initial} \otimes {\rm soft}}\,S^\dagger S   =
  \Tr_{\rm initial}\, \SH^\dagger\SH
  \label{eq:KLN_trace}
\end{equation}
where the first trace is over the degenerate initial states and the unresolved soft
sector.
The $U_I$ operator does not cancel when considering instead a fixed hard initial state,
leaving $S^\dagger S = U_I^\dagger \SH^\dagger \SH U_I \equiv \SH^\dagger \SH + \Delta$.
The $U_I$ operator, expanded to one-soft-emission accuracy, is
\begin{equation}
  U_I\ket{0}
  =
  (1+V)\ket{0}
  +
  \sum_a\int d\Phi_k \,\Rop_a(k)\ket{k,a}
  +\cdots .
  \label{eq:UI_expansion}
\end{equation}
The `real' operator $\Rop_a(k)$ creates one unresolved soft gauge boson of momentum $k$ and adjoint index $a$, and is of order $g_2$.
The `virtual' operator $V$ is the no-emission part of the soft evolution and starts at order $g^2_2$.
At order $g^2_2$ unitarity of $U_I$ links them as
\begin{equation}
  V+V^\dagger
  +
  \sum_a\int d\Phi_k \,\Rop_a^\dagger(k)\Rop_a(k)
  =0.
  \label{eq:unitarity_V}
\end{equation}
This relation fixes the Hermitian part of $V$ and allows to write the  correction $\Delta$ in terms of real effects only\footnote{The anti-Hermitian part of $V$, corresponding to a Coulomb/Glauber phase, contributes to eq.\eq{deltaO_preview} through an additional commutator. It vanishes for the group structure $T^a_1T^a_2$ that arises from virtual soft-boson exchange between the initial states.}:
\begin{equation}
\Delta
  =
  \sum_a\int d\Phi_k
  \left[
    \Rop_a^\dagger\SH^\dagger \SH \Rop_a
    -\frac12\{\Rop_a^\dagger\Rop_a,\SH^\dagger \SH\}
  \right].
  \label{eq:deltaO_preview}
\end{equation}
The Lindblad term on the right side has vanishing trace, reproducing eq.\eq{KLN_trace}.
This is the operator statement that the probability carried by real unresolved one-boson states
is compensated by the no-emission virtual part of the soft evolution.
The real term feeds different gauge charges and the virtual term subtracts probability from the channel present before the soft emission.

\smallskip

In our non-relativistic DM problem, the emission factors $R_a$ are explicitly given by eq.\eq{R_Jpm}.
Its first term involving $T_+^a$ cancels because $\SH^\dagger \SH$ is proportional to the identity inside each total-isospin multiplet.
So the dominant $\beta\to 0$ rates of section~\ref{sec:SU2s} were not affected by IR-enhanced effects.
The remaining  $T_-^a$ terms give  $\beta^2$-suppressed IR-enhanced effects, given by
\beq\Delta
  =
  \calW_-\sum_a
  \left[
  T^a_- \SH^\dagger \SH T^a_- -\frac12\{T^a_- T^a_- ,\SH^\dagger \SH\}
  \right]. \label{eq:final}
\eeq
%Therefore, radiation from the initial state is negligible in the leading $s$-wave.
%and IR divergences cancel when summing over the final state only,
%even when considering a specific DM initial state.

\subsection{Corrections to velocity-suppressed cross sections}\label{sec:pwave}
We now consider the next-to-leading terms in the non-relativistic expansion in terms of small velocities $\beta$.
Squaring the amplitude $\mathscr{A} \sim a_0 + a_1 \beta + a_2 \beta^2+\cdots $
gives two types of cross sections suppressed by $\beta^2$:
genuine $p$-wave effects, arising from $|a_1|^2$,
and velocity-suppressed $s$-waves arising from ${\rm Re}\,(a_0 a_2^*)$.
The two cases need to be distinguished because they receive different Sommerfeld corrections~\cite{1303.0200,1411.6924}.
In a partial waves decomposition for $Y=0$, the two DM particle in the initial state have even $(I-1)/2+\ell+ S$.
So spin $S$ switches as $0\leftrightarrow 1$ when going from $\ell=0$ to $\ell=1$.
The two $\ell$ can be distinguished by decomposing the amplitude into even or odd combinations under
exchange of the DM momenta.
Extending the computation of the tree-level annihilation rates of eq.\eq{SigmaTree} from vanishing to small velocities
gives the following $p$-wave rates
\beq\label{eq:ptree}
\sigma v^{{WW}}_{I=3,\ell=1,S=0} = \frac{\pi \alpha_2^2 (n^2 - 1)}{12 \MDM^2 \gDM}\beta^2  ,\qquad
\displaystyle
 \sigma v^{WW}_{I=1,5,\ell=1,S=1} = \frac{7}{3}\beta^2 \sigma v^{WW}_{I=1,5},  \eeq
and the following $s$-wave rates
\beq\label{eq:fakestree}
 \sigma v^{WW}_{I=1,5,\ell=0,S=0} = -\frac{4}{3}\beta^2 \sigma v^{WW}_{I=1,5} , \qquad
\sigma v^{f\bar f,HH^\dagger }_{I=3,\ell=0,S=1} = -\frac{4}{3}\beta^2\sigma v^{f\bar f,HH^\dagger }_{I=3}.
\eeq
We wrote the velocity-suppressed terms in a redundant notation where $I,\ell, S$ are explicit.
Similar results were derived for $n=3$ in~\cite{1303.0200}.

\smallskip

IR-enhanced Sudakov corrections to such $\beta^2$-suppressed rates can be computed from eq.\eq{final}. 
One can distinguish two different regimes, depending on the energy $E$ of the emitted vector. 
\begin{itemize}
\item If the vector energy $E$ is much smaller than the scale of Sommerfeld corrections, $E \ll \MDM \beta^2$, the total rate can be approximated as $(\text{Sudakov})\cdot (\text{Sommerfeld})_I$. In this case, the Sommerfeld effect is a hard interaction compared to the emission of the soft vector, 
and the usual cancellation between real and virtual IR divergences takes place.\footnote{See however~\cite{2011.04701} and references therein for IR non-cancellation
issues that arise at next-to-next-to-leading order in the presence of massive particles.}

\item
If the vector energy is instead much larger than the Sommerfeld scale, $E \gg \MDM \beta^2$, Sudakov emission is much faster than the Sommerfeld distortion of the initial state, so that the total rate can be approximated as $(\text{Sommerfeld})_{I'} \cdot (\text{Sudakov})$. In this case, there is a non-trivial `commutator' between the Sudakov and Sommerfeld effects, and log-enhanced corrections survive in the total rate. Indeed, the vector emission changes the total isospin $I$ of the DM pair into $I' = I \pm 2$, and the real and virtual amplitudes get different Sommerfeld corrections, which are diagonal in total isospin.

\end{itemize}
%There is a non-trivial `commutator' between the Sudakov and Sommerfeld effects because the Sommerfeld corrections are diagonal in total isospin, while emission of one vector changes $I$ into $I' = I \pm 1$.
Taking into account the weak vector mass $M_W^2 \approx \frac{11}{6} g_2^2 T^2 + \frac12 g_2^2 v^2(T)$ as infrared regulator, see~\cite{hep-ph/0004071},
weak corrections are given by $s$-wave tree cross sections times the regulated
\beq
\calW_- =
%\frac{\alpha_2}{4\pi} \frac{16}{3} \frac{\beta^2}{\epsir} \rightarrow
 \frac{\alpha_2}{4\pi} \frac{4}{3} \beta^2 \ln \frac{E^2_{\rm hard}}{E^2_{\rm Sommerfeld}},
\eeq
where $E_{\rm hard} \sim 2\MDM$ is the maximal energy of the emitted vector, corresponding to the scale of the hard process, and $E_{\rm Sommerfeld}\sim \max (M_W, \MDM \beta^2)$. A more complete treatment is required when the vector is emitted with energy $E\sim \MDM \beta^2$ from inside the Sommerfeld ladder, possibly leading to an extra mild enhancement.
%\DB{These corrections will be of the form $(\text{Sommerfeld})_{I'}\cdot(\text{Sudakov})\cdot(\text{Sommerfeld})_I$. Do we want to say something about why they don't cancel the previous IR-ehnanced term?}  \AS{no}

The IR-enhanced corrections of eq.\eq{final} can be explicitly computed
by writing the hard scattering matrix in terms of the tree-level annihilation matrix $\Gamma$
and by performing its $\SUL$ decomposition, obtaining
%from the emission of a $\SUL$ gauge boson
%can be obtained inserting $\Gamma$ in the form of eq. (\ref{eq:Gamma}) into the general formula (\ref{eq:}).
\beq\Delta\Gamma
  =
  \calW_- \sum_{I'} \Gamma_{I'} \sum_{a}
  \left[
  T^a_- \Pi_{I'} T^a_- -\frac12\{T^a_- T^a_- ,\Pi_{I'}\}
  \right]
 \equiv  \sum_{I}\Delta\Gamma_{I} \Pi_{I} . \label{eq:finalbis}
\eeq
The correction to isospin channel $I$ is
\beq
\Delta\Gamma_{I} = \calW_-\sum_{I'}R_{II'} \Gamma_{I'},
\qquad
R_{II'} = \frac{1}{I}\Tr\left( \,\Pi_I \sum_a \left[
  T^a_- \Pi_{I'} T^a_- -\frac12\{T^a_- T^a_- ,\Pi_{I'}\}
  \right]\right).
\eeq
The second virtual term contributes only to the diagonal part of  $R_{II'}$
\beq
T^2_- \Pi_{I'}= (2T^2_1 + 2 T^2_2 - T^2_+) \Pi_{I'} =  \left(n^2-1-({I'}^2-1)/4\right)\Pi_{I'}.
\eeq
The first real term is~\cite{Edmonds:1955fi}
\beq B_{II'}=\frac{\bar I (4n^2 - \bar{I}^2)}{8},\qquad \bar I = \frac{I+I'}{2}\qquad\hbox{for $|I-I'|=2$}.\eeq
% \AS{can you simplify below?}
%\beq
%B_{II'} = \Tr\left(\Pi_I \sum_a
%  T^a_- \Pi_{I'} T^a_- \right) = \begin{cases}
%\frac{1}{8}(I - 1)(2n - I + 1)(2n + I - 1)& \text{if } I'=I-2, \\
%\frac{1}{8}(I + 1)(2n - I - 1)(2n + I + 1)& \text{if } I'=I+2. \\
%\end{cases}
%\eeq
The real plus virtual sum gives the total matrix
\beq
R_{II'} = \frac{B_{II'}}{I}  - \left[n^2-1- \frac{I'^2-1}{4}\right]\delta_{II'}.
\eeq
This correction to $\Gamma$ can be rewritten in terms of cross sections as 
\beq\delta (\sigma v_{I})=\calW_- \sum_{I'}\hat{R}_{II'}\sigma v_{I'}\qquad
\hat{R} =
\bordermatrix{ & {\scriptstyle I'=1} & {\scriptstyle I'=3} & {\scriptstyle I'=5}  \cr
{\scriptstyle I=1} & 1-n^2 & (n^2-1)/{3} & 0   \cr
{\scriptstyle I=3}& n^2-1 & 3-n^2 & {2(n^2-4)}/{5}   \cr
{\scriptstyle I=5}& 0 & {2(n^2-4)}/{3} & 7-n^2   \cr
{\scriptstyle I=7}& 0 & 0 & {3(n^2-9)}/{5}
}.\label{eq:IRcorbeta}
\eeq
{\begin{figure}[t]
$$\includegraphics[width=0.8\textwidth]{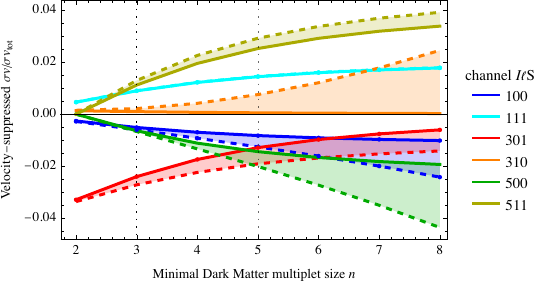}$$
\begin{center}
\caption{\em 
The continuous curves show the tree level values of the velocity-suppressed cross sections divided by the total rate
obtained by naively summing the $s$-wave cross sections in eq.\eq{SigmaTree}.
The dashed curves include the IR-enhanced EW corrections.
\label{fig:EWcorrectionMDMvsuppressedAbs}}
\end{center}
\end{figure}}%
where we explicitly listed the relevant entries of the rescaled matrix $\hat{R}$.
This correction gets larger for larger multiplets, in addition to being IR-enhanced.
The vanishing of the IR-enhanced corrections when summing over all components corresponds
to the vanishing sum of the entries in each column, $\sum_I \hat{R}_{II'}=0$.
Explicitly, eq.\eq{IRcorbeta} is
\begin{eqnsystem}{sys:SS}
\delta \sigma v^{WW}_{I=1,\ell=0,S=0} &=& \sigma v^{WW}_{I=1}(1-n^2) \mathcal{L}_-, \label{eq:WWvirtual1} \\ 
\delta \sigma v^{{WW}}_{I=3,\ell=1,S=0} &=& \bigg[\sigma v^{WW}_{I=1}(n^2-1) +\sigma v^{WW}_{I=5} \frac{2(n^2-4)}{5}\bigg]  \mathcal{L}_- ,\\
\delta\sigma v^{f\bar f W,HH^\dagger W}_{I=1, \ell=1 , S=1} &=& \sigma v^{f\bar f,HH^\dagger}_{I=3} \frac{n^2-1}{3} \mathcal{L}_-, \label{eq:HHreal1}\\
\delta\sigma v^{f\bar f,HH^\dagger}_{I=3,\ell=0,S=1} &=& \sigma v^{f\bar f,HH^\dagger}_{I=3}\left[-\frac{n^2-1}{3} -\frac{2(n^2-4)}{3}\right] \mathcal{L}_- ,
\label{eq:HHvirtual} \\
\delta \sigma v^{WW}_{I=5,\ell=0,S=0} &=& \sigma v^{WW}_{I=5}(7-n^2) \mathcal{L}_-,  \label{eq:WWvirtual5}\\
\delta\sigma v^{f\bar f W,HH^\dagger W}_{I=5,\ell=1,S=1} &=& \sigma v^{f\bar f,HH^\dagger}_{I=3}\frac{2(n^2-4)}{3} \mathcal{L}_- ,\label{eq:HHreal5} \\
\delta \sigma v^{{WW}}_{I=7,\ell=1,S=0} &=&\sigma v^{WW}_{I=5} \frac{3(n^2-9)}{5} \mathcal{L}_-  .
\end{eqnsystem}
where virtual effects contribute to eq.\eq{WWvirtual1},\eq{HHvirtual},\eq{WWvirtual5} and real emission to the others.
%$\sigma_0$ is the tree-level cross-section and (eq. (3.2) in \cite{1305.6391}\AS{write in terms of our $\sigma v_{I=3}$?}),
Fig.\fig{EWcorrectionMDMvsuppressedAbs} compares the tree-level values of eq.\eq{ptree},\eq{fakestree} with the $\sigma v$ obtained after adding the IR-enhanced weak corrections in eq.~(\ref{sys:SS}).
We assumed $\med{\beta^2} =2T/3\MDM$ around decoupling at $T \approx\MDM/25$. 
These relatively large EW corrections cancel in the total rate if computed as a flat sum over all isospin channels, omitting Sommerfeld factors.
The Sommerfeld factors disrupt the cancellation and are
sizeable if $M_W \lesssim \alpha_{\rm eff} M$ and $\beta\lesssim \alpha_{\rm eff}$, where $\alpha_{\rm eff}\sim n^2 \alpha_2$.
As a result, the velocity-suppressed cross-sections receive a relatively significant weak correction.

\section{Results}\label{sec:235}
In the previous sections we computed the one-loop weak corrections to Minimal Dark Matter annihilation cross sections. Thermal corrections can be neglected, since freeze-out occurs at relatively low temperatures, $T \lesssim M_X/25$. We now apply our results to the most motivated multiplets. 
The `thermal' masses that reproduce the observed dark matter abundance lie in different regimes, so different approximations are appropriate in each case.

%In addition to quantum corrections $K_0$, thermal effects give further corrections of the schematic form
%\beq \sigma v|_{\rm NLO} = \sigma v \bigg[ 1 + K_0 \frac{g^2}{(4\pi)^2} + K_T g^2 \frac{T^2}{\MDM^2} + \cdots\bigg].\eeq
%For example, the mass of slow DM particle acquires the thermal correction $\delta \MDM^2 = g_2^2 C_n /6$ with $C_n = (n^2-1)/4$,
%that contributes as $K_T \approx -g_2^2 C_n T^2/6$.
%These thermal corrections are negligible, at the per-mille level, because
%freeze-out occurs at temperature $T \lesssim M_X/25$.

\subsection{Higgsino-like doublet}\label{sec:2}
We first consider DM as an Higgsino-like $n=2$ doublet with hypercharge $Y=1/2$.
Its `thermal' mass is $\MDM \sim \TeV$, low enough that Sommerfeld corrections to the annihilation cross sections are small and bound-state effects are absent.
It is therefore sufficient to compute the one-loop correction to the total $s$-wave cross section, averaged over all DM components.
The tree level result is
\beq \sigma v|_{\rm tree}=\frac{\pi  \left(81 \alpha _2^2+12 \alpha _2 \alpha _Y+43 \alpha _Y^2\right)}{64 \MDM^2} (1+\delta_ M)\eeq
where all couplings are renormalized at $\bar\mu=2\MDM$ and $\MDM$ is the pole mass.
In the first term all SM particles are approximated as massless.
Since $\MDM$ is only mildly above the weak scale, we also computed the effect of SM particle masses,
at leading order in $M^2_{W,Z,t,h}/ \MDM^2$, obtaining the correction
$\delta_M \approx -0.12 \% (\TeV/\MDM)^2$.
Such correction is analytically given by
$\delta_M \approx (47 M_W^2-3M_h^2-18 M_t^2)/216\MDM^2$
in the limit $g_Y=0$ where $M_W =M_Z$.\footnote{Keeping $M_W\neq M_Z$, the full
mass correction to the total cross section at leading order in $1/\MDM$ is
\beq
\delta_M \sigma v|_{\rm tree} =
-\frac{3 \left(M_h^2+6 M_t^2\right) (4 M_W^4+M_Z^4-2 M_W^2 M_Z^2)}{512 \pi  \MDM^4 V^4}+
\frac{300 M_W^2   M_Z^2(M_W^2-M_Z^2)+320 M_W^6+103 M_Z^6}{1536 \pi  \MDM^4 V^4}
   \eeq
where $V=246\GeV$.}
This expression can be adapted to also include thermal mass corrections,
of order $g^2 T^2/\MDM^2$ at sub-$\%$ level.
%taking into account that all particles acquire thermal masses.
The one loop correction is IR finite:
%\frac{81\pi \alpha_2^2(2M)}{64 \MDM^2} (1+\delta_Y)
%\qquad
%\delta_Y= \frac{4 \alpha_Y}{27 \alpha_2}+\frac{43 \alpha_Y^2}{81 \alpha_2^2} \approx 0.09 \eeq
\beq\begin{split}
 \sigma v|_{\rm loop}&=\frac{1}{16 \MDM^2}\bigg[\alpha _2^3 \left(\frac{2167}{48}+\frac{57 \pi ^2}{32}+25 \ln2\right)+
 \alpha _2^2 \left(\frac{27}{16} \left(8 \alpha _3-\alpha _t \right)- \frac{382+45  \pi ^2}{96} \alpha _Y\right)+\cr
 &+\frac{1}{16} (88 \alpha _3-17 \alpha _t) \alpha _Y^2+
 \frac{171 \pi ^2-5950}{96}  \alpha _2 \alpha  _Y^2+
 \frac{57 \pi ^2-23374}{288}  \alpha _Y^3\bigg] \approx 0.048 \sigma v|_{\rm tree}.
\end{split} \eeq
We here also included hyper-charge and top Yukawa corrections
in addition to weak and strong interactions.
Higgs quartic corrections vanish, at one loop.

Our computation takes into account Sommerfeld, one loop and mass corrections to the $s$-wave cross section,
while approximating the $p$-wave cross section with its tree-level value.
The `thermal' value of the mass increases  from $\MDM=1024\GeV$ to $\MDM=1046\GeV$ in view of loop and mass corrections.
It's difficult to compare with loop computations performed in~\cite{0710.1821,0910.3293,1210.7928,1411.6930,1912.02034}
in the vast parameter space of supersymmetric models.
Possibly small non-minimal effects (such as the existence of the bino) are needed to make the Higgsino compatible with direct detection bounds.

\subsection{Wino-like triplet}
We next consider the wino-like fermion 3-plet with $Y=0$.
Our numerical computation includes EW and QCD corrections, bound states, vector thermal masses,
but it does not include higher-order corrections to weak potentials and Sommerfeld corrections to $p$-wave processes.

A precise computation in components of Sommerfeld corrections in $s$ and $p$ waves
finds that $\Omega_{\rm DM} h^2 = 0.1205$ is reproduced for
$\MDM=2842\GeV$~\cite{2009.00640}, using tree level expressions with $\alpha_2$ renormalized at $\bar\mu=2\MDM$,
% said at page 21
and the tree level value of $\MDM$.
See also~\cite{2107.09688}.
Precisely defining the DM mass $\MDM$ as the pole mass, its `thermal' value
 is increased by $0.6\%$ by QCD corrections,
and decreased by $0.2\%$ by $\alpha_{2,Y,t}$ EW corrections to the $s$-wave cross sections.
Bound state formation (not included in~\cite{2009.00640}) increases $\MDM$ by $1\%$.
Adding bound state, EW and QCD corrections modifies the mass claimed by~\cite{2009.00640} into
\beq \MDM =2881\GeV\qquad\hbox{(wino-like fermion 3-plet with $Y=0$)}.\eeq
The thermal relic abundance that matches the observed DM abundance is well approximated by the $\SUL$-invariant limit.

\subsection{Quintuplet}
We next consider the stable fermion 5-plet with $Y=0$.
Including Sommerfeld-corrected annihilations,
EW plus QCD corrections reduce the thermal mass by $4\%$,
bringing it to $\MDM\approx 8.2\TeV$.
However, bound state formation gives an important correction, giving $\MDM \approx 14\TeV$~\cite{1702.01141}.
We have not computed EW corrections to bound state formation, so we cannot provide a more precise prediction for $\MDM$.

\section{Conclusions}\label{sec:concl}
We computed the one-loop electroweak corrections to the annihilation cross sections of fermionic Minimal Dark Matter multiplets, 
with the aim of improving predictions for the thermal relic abundance and the DM mass.

The dominant $s$-wave annihilation rates are free from Sudakov-enhanced electroweak logarithms. 
Although real and virtual weak corrections separately contain infrared singularities, these singularities cancel in the inclusive combinations relevant for the relic abundance.
This cancellation persists even in the presence of Sommerfeld effects, 
because in the non-relativistic limit soft vectors couples at leading order to the total weak charge of the two-particle state.
Initial-state soft radiation is therefore absent, and the KLN cancellation proceeds channel by channel without interference from Sommerfeld factors.
Relatively large corrections arise for large multiplets $n$.
Fig.\fig{EWcorrectionMDM} illustrates the numerical relevance of such corrections,
eq.\eq{sigmaLNO} provides expressions in the $\SU(2)_L$-invariant limit, and section~\ref{sec:components} shows their decomposition in components.

The situation is different for velocity-suppressed annihilation rates.
There is now a soft-emission current that connects different total-isospin representations,
so that, in view of [Sudakov, Sommerfeld] non-commutation effects,
Sudakov-enhanced corrections no longer cancel channel by channel once the Sommerfeld factors weight the different isospin channels differently.
Residual IR-enhanced corrections affect $p$-wave and velocity-suppressed $s$-wave cross sections.
Results are given in eq.~(\ref{sys:SS}) and illustrated in fig.\fig{EWcorrectionMDMvsuppressedAbs}.

\smallskip

We then applied these one-loop results to the most motivated fermionic multiplets.
\begin{enumerate}
\item[2)] The Higgsino-like doublet annihilation cross section receives a $5\%$ correction.

\item[3)]Electroweak loop effects grow with the multiplet size as illustrated in fig.\fig{EWcorrectionMDM},
but accidentally happen to be small for the Wino-like triplet.

\item[5)] For the stable quintuplet, loop corrections effects are at the $10\%$ level. 
However, a complete NLO prediction is not yet possible, since electroweak corrections to bound-state formation, which gives a large contribution, remain to be computed.
\end{enumerate}

%The thermal correction to the elctron mass is $\delta m_e^2 = e^2 T^2/6$. So in the non-abelian case this becomes $\delta M^2 = g^2 C_R T^2/6$.
%For the fundamental of SU($N$) $C_N = (N^2-1)/2N$, and $C_n = (n^2-1)/4$ for a $n$-dimensional representation of $\SU(2)$.

\footnotesize

\section*{Acknowledgments}
P.P.G. is supported by the Ram\'on y Cajal grant~RYC2022-038517-I funded by MCIN/AEI/10.13039/501100011033 and by FSE+, and by the Spanish Research Agency (Agencia Estatal de Investigaci\'on) through the grant IFT Centro de Excelencia Severo Ochoa~No~CEX2020-001007-S.
A.S.\ thanks Martin Beneke, Paolo Ciafaloni, Claude, Denis Comelli, Chat GPT, Diego Redigolo for clarifications.

\appendix\small

\section{Computing one loop weak corrections}\label{sec:appA}

\subsection{Renormalization}
Here we discuss details of the calculation of the one-loop contributions presented in the main text. We compute the relevant amplitudes in the
Feynman - 't Hooft gauge using FeynArts \cite{Hahn:2000kx} and FeynCalc~\cite{Shtabovenko:2020gxv}. The FeynArts model file was created using FeynRules \cite{Alloul:2013bka}.
The UV and IR divergences were treated using dimensional regularization in $d=4-2\epsilon$ dimensions.  To renormalize the theory we define opportune counter-terms for the weak gauge coupling and the mass of the DM multiplet.

Before proceeding, it is worth clarifying our notation for the weak gauge coupling $\alpha_2=g_2^2/4\pi$, which we define as its SM value in the $\overline{\text{MS}}$ scheme. This differs\footnote{This difference does not affect the discussion in this appendix, as it is of higher order. However it enters in the final results presented in the main text.} from the MDM weak gauge coupling $\alpha_2^{\rm MDM}$ by the additional running up to the DM multiplet scale.
The two definitions of weak gauge couplings can be matched at the renormalization scale $\bar\mu=\MDM$ as
\beq
\alpha_2(\bar\mu) \equiv \alpha_2^{\rm SM}(\bar\mu) = \alpha_2^{\rm MDM}(\bar\mu)\left[ 1 + \frac{\alpha_2^{\rm MDM}(\bar\mu)}{4\pi}(\beta_2-\beta_2^{\rm SM}) \ln \left( \frac{\bar\mu^2}{\MDM^2} \right) \right],
\eeq
where
\beq
\beta_2^{\rm SM}= \left[\frac{11}{3}C(3) - \left(\frac{2}{3}n_L + \frac{1}{3}\right)T(2)\right],
\eeq
$C$ is the Casimir of the gauge symmetry and $T$ the Dynkin index of the representation.

The counter term for the gauge coupling is:
\beq
\delta Z_g^{\msbar} = -\frac{\alpha_2}{8\pi}\beta_2 \frac{1}{\epsilon} =  -\frac{\alpha_2}{8\pi}\left[\beta_2^{\rm SM} - \frac{\gDM}{3n} T(n)\right] \frac{1}{\epsilon}.
\label{eq:couplingct}
\eeq
For the renormalization of the DM multiplet we first define the two-point function:
\begin{align}
\Sigma(p)=(\slashed{p} \Sigma_{p}+\Sigma_{m}),
\end{align}
where $p$ is the momentum. The On Shell (OS) counter-term is then defined from the relation
\begin{align}
M_{\DM}=M_{0,\DM}-\delta M^{\OS}_{\DM}\,
\end{align}
between the bare $M_{0,\DM}$ and physical mass  and $M_{\DM}$:
\begin{align}
\delta M^{\OS}_{\DM}= \MDM \Sigma_{p}+ \Sigma_{m}\,
\end{align}
which for us it gives
\beq
\delta M^{\OS}_{\DM} =  \frac{\alpha_2}{4\pi} C(n) \MDM\left[\frac{3}{\epsilon} +3  \ln\left(\frac{\bar\mu^2}{\MDM^2}\right) + 4 \right].
\eeq
Finally, the OS wave-function renormalizations are
\beq
\delta Z^{\OS}_W =-\frac{d\Pi_{WW}}{dp^2}|_{p^2=0}= -\frac{\alpha_2}{4\pi}\left[\frac{\gDM}{3n}T(n)\left(  \frac{1}{\epsilon} + \ln\left(\frac{\bar\mu^2}{\MDM^2}\right)  \right)\right]\, ,
\eeq
\beq
\delta Z^{\OS}_{\DM} = -\Sigma_{p}-2\frac{d}{dp^2}\left(\MDM^2 \Sigma_{p}+ \MDM \Sigma_{m} \right) =-\frac{\alpha_2}{4\pi}C(n)\left[\frac{3}{\epsilon}  +3 \ln\left(\frac{\bar\mu^2}{\MDM^2}\right) + 4 \right]\, ,
\eeq
where $\Pi_{WW}$ is the $\SU(2)_L$ gauge vectors 2-point function. The wave-function renormalizations of other fields involved in the calculation are zero.

\subsection{Virtual corrections to $\DM\bar\DM\to f\bar{f},HH^\dagger$}
We compute the virtual contributions $\delta^\Fin_{\rm vir}$ to DM annihilations into $\SUL$ doublets  $\Fin=f\bar{f},H H^\dagger$, by calculating the interference of the tree-level result with the one-loop amplitudes. Since only the amplitudes with isospin $I=3$ are non-zero at the tree-level, contributions to other isospins drop in the interference. After summing up all the virtual corrections, including the counter-term contributions as defined in the previous section, we find:
\begin{align}
\delta^{f\bar f}_{{\rm vir},I=3} &= \frac{\alpha_2}{\pi}\left[\frac{\pi^2(n^2-5)}{8\beta}-\frac{3}{4\epsilon^2} - \frac{17+6 {\cal L}}{8\epsilon} -\frac{\gDM(n^2-1)(8+3{\cal L}+6\ln2)}{216}\right.\nonumber\\&+\left. \frac{(-144 n^2 - 40 n_L - 6 {\cal L} (-35 + 9 {\cal L} + 4 n_L) - 9 \pi^2 +
  64 (20 + 3\ln2))}{144}\right]\nonumber \\
  &-(\frac{\alpha_3}{\pi}+\frac{\alpha_Y}{12\pi})\left[\frac{1}{\epsilon^2}+\frac{3+2 {\cal L}}{2\epsilon}+  \frac{7 \pi^2 -6 (8 + {\cal L} (3 + {\cal L}))}{12}\right]+\frac{\alpha_t}{\pi}\left[\frac{1}{16\epsilon}+\frac{3+{\cal L}}{16}\right],\\
\delta^{H H^\dagger}_{{\rm vir},I=3} &=\left.\delta^{f\bar f}_{{\rm vir},I=3}\right|_{\alpha_3=0}-\frac{\alpha_2}{\pi}\left[\frac{3}{8\epsilon}+\frac{3}{8}{\cal L} +1\right]+\frac{\alpha_t}{\pi}\left[\frac{23}{16\epsilon}+\frac{45+23{\cal L}}{16}\right]\nonumber \\
  &-\frac{\alpha_Y}{12\pi}\left[\frac{2}{\epsilon^2}+\frac{9+4 {\cal L}}{2\epsilon}+  \frac{48 + 27 {\cal L} + 6 {\cal L}^2 - 7 \pi^2}{6}\right].
\end{align}
The terms proportional to $1/\beta$, where $\beta=\sqrt{s/4\MDM^2-1}$ is the DM velocity in the center of mass frame,
describe weak potential effects accounted, after resummation, by Sommerfeld factors.

\subsection{Virtual corrections to $\DM\bar\DM\to WW$}
At tree-level, in the $s$-wave limit, only the amplitudes in  the  $I=1$ and $I=5$ channels are non-zero.
The diagrams with a $W$ in the $s$ channel do not contribute to the virtual corrections since the intermediate $W$ forces $I=3$:
\begin{flalign}
\delta^{WW}_{{\rm vir},I=1} = & \frac{\alpha_2}{\pi} \left[\frac{\pi^2(n^2 - 1)}{8\beta} - \frac{2}{\epsilon^2} - \frac{43 + 24 {\cal L} - 2 n_L}{12\epsilon}\right. \nonumber \\+&\left.\frac{156 - 48 {\cal L}^2 + 29 \pi^2 + 3 n^2 (-20 + \pi^2)}{48}-\frac{\gDM(n^2-1){\cal L}+2\ln2}{72} \right],\\
\delta^{WW}_{{\rm vir},I=5} = & \delta^{WW}_{{\rm vir},I=1}- 3\frac{\alpha_2}{\pi} \left[\frac{\pi^2}{2\beta}+\frac{1}{\epsilon}+{\cal L}+\frac{\pi^2}{4}-2\right].
\end{flalign}

\subsection{Real corrections}
We present our real emission results in the $s$-wave limit  in terms of the factor $\delta^\Fin_{\rm real}$.
\begin{align}
\delta^{WW}_{1,\rm real} &=\frac{\alpha_2}{\pi}\left[ \frac{2}{\epsilon^2} + \frac{2{\cal L} + \frac{11}{3}}{\epsilon} + {\cal L}^2 + \frac{11{\cal L}}{3} - \frac{7\pi^2}{4} + \frac{181}{9} \right],
\\
\delta^{WW}_{5,\rm real} &=\delta^{WW}_{1,\rm real}+3\frac{\alpha_2}{\pi}\left[ \frac{1}{\epsilon} +{\cal L}+3-\frac{\pi^2}{12} \right],
\\
\delta^{f\bar f}_{3,\rm real} &=\frac{\alpha_2}{\pi}\left[ \frac{3}{4\epsilon^2} + \frac{6{\cal L} + 17}{8\epsilon} + \frac{1}{48}(18{\cal L}^2 + 102{\cal L} - 21\pi^2 + 347) \right]\nonumber \\
  &+(\frac{\alpha_3}{\pi}+\frac{\alpha_Y}{12\pi})\left[\frac{1}{\epsilon^2}+\frac{3+2 {\cal L}}{2\epsilon}+  \frac{57 + 6 {\cal L} (3 + {\cal L}) - 7 \pi^2}{12}\right]-\frac{\alpha_t}{\pi}\left[\frac{1}{16\epsilon}+\frac{8+3{\cal L}}{48}\right],
\\
\delta^{H H^\dagger }_{3,\rm real} &=\left.\delta^{f\bar f}_{3,\rm real}\right|_{\alpha_3=0}+ \frac{3}{8}\frac{\alpha_2}{\pi}\left[ \frac{1}{\epsilon}+{\cal L}+\frac{9}{2}\right]-\frac{\alpha_t}{\pi}\left[\frac{23}{16\epsilon}+\frac{268+69{\cal L}}{48}\right]\nonumber\\&+\frac{\alpha_Y}{12\pi}\left[\frac{2}{\epsilon^2}+\frac{9+4 {\cal L}}{2\epsilon}+  \frac{195 + 54 {\cal L} + 12 {\cal L}^2 - 14 \pi^2}{12}\right],
\\
\delta^{f\bar f}_{1,\rm real} &=\delta^{f\bar f}_{5,\rm real} =-\frac{\alpha_2}{\pi}\left[ \frac{n_L}{6\epsilon} +\frac{(3{\cal L} + 8)n_L}{18} \right],\label{eq:realff1}
\\
\delta^{H H^\dagger }_{1,\rm real} &=\delta^{H H^\dagger }_{5,\rm real}= -\frac{\alpha_2}{\pi}\left[ \frac{1}{12\epsilon} + \frac{(3{\cal L} + 11)}{36} \right].
\label{eq:realhh1}
\end{align}
The IR divergences in eq.s~(\ref{eq:realff1}-\ref{eq:realhh1}) do not originate from a soft or collinear emission of $W$, but rather from the propagator of the mediating $W$ going on shell. These contributions cancel the remaining IR-divergences in the virtual correction in the respective $WW$ annihilation channels arising from $\delta Z^{\OS}_W$. Therefore, we include these corrections in the total correction $\delta^{WW}_{I=1}$ and $\delta^{WW}_{I=5}$ given by eq.~(\ref{sys:delta}).

\footnotesize

\begingroup
\renewcommand{\baselinestretch}{0.95}\selectfont

\endgroup

\end{document}